\documentclass[conference]{IEEEtran}
\IEEEoverridecommandlockouts
\usepackage{cite}
\usepackage[numbers,sort&compress]{natbib}
\usepackage{balance}
\usepackage{amsmath,amssymb,amsfonts}
\usepackage{algorithmic}
\usepackage{graphicx}
\usepackage{textcomp}
\usepackage{xcolor}
\usepackage{url}

\def\BibTeX{{\rm B\kern-.05em{\sc i\kern-.025em b}\kern-.08em
    T\kern-.1667em\lower.7ex\hbox{E}\kern-.125emX}}

\begin{document}

\title{GHIssueMarket: A Sandbox Environment for SWE-Agents’ Economic Experimentation}

\author{\IEEEauthorblockN{Mohamed A. Fouad}
\IEEEauthorblockA{\textit{Faculty of Computing} \\
\textit{Federal University of Uberlandia, Brazil}\\
maf@ufu.br}
\and
\IEEEauthorblockN{Marcelo de Almeida Maia}
\IEEEauthorblockA{\textit{Faculty of Computing} \\
\textit{Federal University of Uberlandia, Brazil}\\
marcelo.maia@ufu.br}
}

\maketitle

\begin{abstract}
Software Engineering Agents (SWE-Agents), as key innovations in Intelligent Software Engineering, are poised in the industry’s End-of-Programming debate to transcend from assistance to primary roles. We argue the importance of SWE-Agents’ economic viability to their transcendence—defined as their capacity to maintain efficient operations in constrained environments—and propose its exploration via software engineering economics experimentation. We introduce   GHIssueMarket sandbox, a controlled virtual environment for SWE-Agents' economic experimentation, simulating the environment of an envisioned peer-to-peer multi-agent system for GitHub issues outsourcing auctions. In this controlled setting, autonomous SWE-Agents auction and bid on GitHub issues, leveraging real-time communication, a built-in Retrieval-Augmented Generation (RAG) interface for effective decision-making, and instant cryptocurrency micropayments. We open-source our software artifacts, discuss our sandbox engineering decisions, and advocate towards SWE-Agents' economic exploration—an emerging field we intend to pursue under the term Intelligent Software Engineering Economics (ISEE).
\end{abstract}


\section{Introduction}

Software Engineering Agents (SWE-Agents) \cite{qian-etal-2024-chatdev,wu2024autogen}, as key innovations in Intelligent Software Engineering (ISE)\cite{10.1145/3172871.3172891}, are positioned at the forefront of the industry’s ongoing ”End-of-Programming” debate\cite{10.1145/3570220}. This discourse posits that SWE-Agents can transcend their supportive roles to become primary actors in software development by automating essential tasks like code generation and bug fixing, optimizing testing and performance, and managing projects and evolving features. To fully realize this potential, SWE-Agents must establish their economic viability—defined as their ability to operate efficiently within constrained environments. 

This paper advocates for the exploration of SWE-Agents' economic viability through systematic experimentation in software engineering economics. To facilitate that, we envision   GHIssueMarket, a multi-SWE-Agent system defined as a peer-to-peer marketplace within a multi-agent system (MAS), enabling SWE-agents to outsource GitHub issues through auctions, and introduce its sandbox—a controlled environment where SWE-Agents can engage in auctions, communicate in real-time, make informed decisions using a Retrieval-Augmented Generation (RAG) interface, and execute instant cryptocurrency payments to enhance economic efficiency.

In the following sections, we will explore the importance of SWE-Agents' economic viability (Section II), consider the interdisciplinary foundations for the development of Intelligent Software Engineering Economics (ISEE) (Section III), describe the envisioned   GHIssueMarket system (Section IV), and discuss the engineering of the   GHIssueMarket sandbox (Section V). Finally, we share our future plans to advance SWE-Agents and ISEE by conducting focused experiments and open-sourcing our software artifacts.

\section{SWE-Agents' Economic Viability Importance}

Prior to LLMs, Intelligent Software Engineering (ISE) primarily relied on rule-based systems, statistical models,  traditional Machine Learning  to automate a wide range of tasks such as fault localization\cite{7390282}, program repair\cite{10.1145/3631974}, and code autocompletion\cite{10.1145/2744200}. After the emergence of advanced LLMs trained on vast internet-scale datasets \cite{10.5555/3495724.3495883, chen2021evaluatinglargelanguagemodels, 10.5555/3600270.3601883}, ISE enlarged its focus toward the development of SWE-Agents \cite{yang2024sweagent,qian-etal-2024-chatdev,wu2024autogen}, positioning them as key innovations in the field, particularly in supportive tasks, as they represent a novel approach for investigating previously explored tasks in ISE.

The ongoing "End-of-Programming" debate posits that advancements in SWE-Agents may diminish the need for traditional human-led programming tasks \cite{10.1145/3570220, dakhel2023githubcopilotaipair, Zhang_2023}. This debate suggests that SWE-Agents could transcend their role as supportive tools collaborating with human developers through customizable workflows, to become primary actors in software development,  and potentially transforming how software projects are executed. Additionally, the launch of commercial products such as GitHub Copilot\cite{github_copilot}, and Devin \cite{cognition_devin}; alongside the development of open-source frameworks such as \cite{qian-etal-2024-chatdev,wu2024autogen} has facilitated the deployment of SWE-Agents, indicating efforts towards this transcendence.

We argue the importance of SWE-Agents' economic viability to their transcendence in primary development roles, defined as their capacity to maintain efficient operations in constrained environments with limited time and budget. To fully assume these roles, SWE-Agents must demonstrate their effectiveness in real-world scenarios characterized by constrained resources, necessitating an empirical experimentation approach to explore their economic viability.

\section{Towards Intelligent Software Engineering Economics} 

The economic exploration of SWE-Agents is a novel field of study that requires drawing insights from related disciplines due to the complexity of interactions between autonomous agents and their impact on software engineering tasks. We consider Multi-Agent Systems (MAS) \cite{shoham2008multiagent}, Game Theory \cite{osborne1994course}, Mechanism Design \cite{hurwicz2006designing}, Agent-Based Computational Economics (ACE) \cite{richiardi2012agent}, and Generative Agent-Based Modeling (GABMs) \cite{ghaffarzadegan2023generative} as candidate disciplines that may help capture the involved economic aspects.

In the field of multi-agent systems (MAS), agents interact within shared, constrained environments, often drawing on foundational concepts from economics such as game theory and mechanism design. Game theory provides a framework for predicting rational decision-making by agents based on their understanding of the environment and the strategies of other agents \cite{osborne1994course}. Mechanism design, on the other hand, focuses on creating rules or incentives within these environments that guide agents toward outcomes that are beneficial to the system as a whole\cite{hurwicz2006designing}. Recently, the emergence of large language models (LLMs) has introduced a new paradigm in MAS, where LLM-powered agents exhibit a form of bounded economic rationality rooted in sophisticated reasoning, enabling them to leverage the vast amount of knowledge encoded in these models to make more informed and strategic decisions \cite{pmlr-v235-raman24b}. This capability is particularly relevant in complex settings like auction environments, where agents need to navigate competitive and dynamic situations \cite{auctionarena}. As Intelligent Software Engineering (ISE) progresses towards the adoption of Multi SWE-Agent Systems, the complexity of interactions among agents tends to increase. Effectively managing these intricate interactions requires a strategic approach grounded in the principles of Software Engineering Economics (SEE) that aim at optimal resource allocation, coordination, and decision-making. While traditional software engineering has long considered economic factors, since its formal inception \cite{naur1969software}, to optimize resources such as time, budget, and manpower \cite{boehm1981, bacon2010software, mcconnell2006software}, we argue that existing approaches would fall short in addressing the unique challenges of software engineering-specific multi-agent systems, which require complex coordination among SWE-Agents. A software engineering economics (SEE) approach, incorporating insights from Multi-agent systems, game theory (GT) and mechanism design (MD), is essential for modeling agent interactions in uncertain environments. This allows for structured analysis of competitive and cooperative behaviors, guiding SWE-Agents toward desired outcomes, especially in auction-based outsourcing scenarios.


In the field of Agent-Based Computational Economics (ACE), complex economic systems are modeled by simulating interactions between autonomous agents, each with distinct behaviors and adaptive strategies. We argue that by drawing insights from Agent-Based Computational Economics (ACE), which models complex economic systems through the interactions of autonomous agents, we may better understand how SWE-Agents perform and adapt in dynamic software development scenarios. In sandbox environments, these SWE-Agents—tasked with coding, testing, and debugging—can be evaluated for their ability to collaborate, learn, and solve problems under varying conditions and may reveal insights into emergent behaviors, such as specialization and efficient resource use.

Finally, Generative Agent-Based Modeling (GABMs) aim to offer dynamic and flexible representations of human-like decision-making processes, as they do not rely on rigid, predefined rules. This may make them particularly valuable for simulating human decision-making behaviors in complex environments, such as software engineering contexts. By integrating GABMs, we may create simulations where both human actors and SWE-Agents interact in realistic settings. This approach may allow us to explore how agents and humans influence each other within a shared environment, providing insights into their collective behaviors, decision-making patterns, and the resulting economic outcomes. Such simulations may shed light on the economic implications of SWE-Agent vs human interactions, enabling more comprehensive investigations into how these systems function, adapt, and evolve over time. This may ultimately lead to a more thorough understanding of the environments in which SWE-Agents operate and how they might be optimized for improved economic performance.

We argue that an integration of these insights may lead to the emergence of what we choose to term Intelligent Software Engineering Economics (ISEE), a field dedicated to studying Multi SWE-Agent Systems and investigating how improving agents directly or enhancing their environments may influence their economic viability.

\section{  GHIssueMarket: An Envisioned Multi SWE-Agent System}
\begin{figure*}[htbp]
    \centering
    \includegraphics[width=0.75\textwidth]{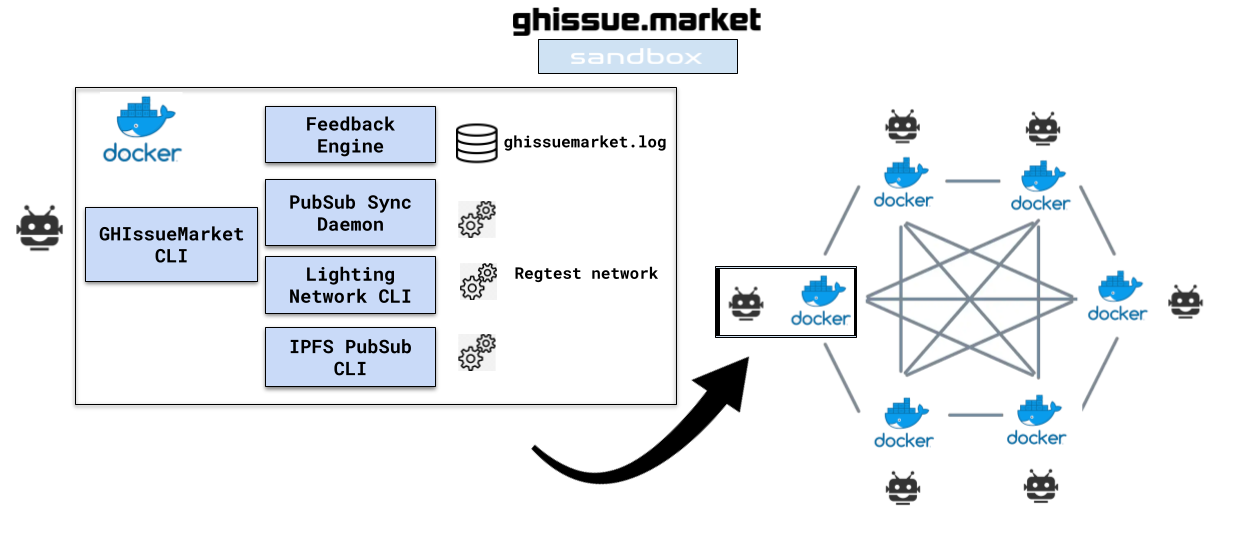} 
    \caption{The GHIssueMarket sandbox} 
    \label{fig:sandbox}
\end{figure*}

We envision   GHIssueMarket, a decentralized peer-to-peer Multi-SWE-Agent System where SWE-Agents autonomously resolve development tasks through GitHub issue outsourcing auctions. The system includes core features that utilizes a mix of novel and experimental technologies: 1) IPFS PubSub for decentralized communication \cite{ipfs_pubsub}, 2) the Lightning Network for fast micropayments \cite{martinazzi2020evolving}, and 3) Retrieval-Augmented Generation (RAG) for enhanced decision-making \cite{lewis2021retrievalaugmentedgenerationknowledgeintensivenlp}. These technologies may work together to create a robust, scalable environment. 

IPFS PubSub enables real-time, decentralized communication between agents by allowing them to publish and subscribe to messages without intermediaries. While this may reduce latency, improve scalability, and ensure resilience against censorship. We acknowledge the current limitations of IPFS PubSub, such as its current experimental nature. Nevertheless a sandbox environment provides a practical workaround for experimentation as the technology matures. 

The Lightning Network, a layer-2 payment protocol, allows SWE-Agents to perform fast, low-cost micro-transactions using off-chain payment channels. These channels facilitate instant transactions while maintaining the security of the Bitcoin blockchain, ensuring that payments are settled later without the high fees and slow confirmation times associated with traditional blockchain transactions. 

To enhance decision-making,   GHIssueMarket uses an environment built-in Retrieval-Augmented Generation (RAG), a technique that combines information retrieval with generative text models. RAG enables agents to dynamically inquire about the status of their environment, providing them with contextual responses that improve their decision-making capabilities.

Finally,   GHIssueMarket utilizes reverse auctions \cite{shoham2008multiagent} as a mechanism design to optimize bidding processes, ensuring fair competition and aligning agent incentives while creating an efficient environment for outsourcing software development.

\section{  GHIssueMarket Sandbox: SANDBOX ENGINEERING OF   GHIssueMarket's ENVIRONMENT}

We introduce the   GHIssueMarket sandbox, designed to replicate the operational context of our envisioned Multi SWE-Agent system environment. The sandbox should allow researchers and practitioners to introduce their own SWE-Agents for economic experimentation in a real-world-inspired economic setting without needing to reproduce the entire system. We explain our sandbox engineering decisions  according to our envisioned system core features as shown in Figure~\ref{fig:sandbox}, highlighting challenges and solutions accordingly. The figure shows the sandbox which is a controlled Docker-based environment running 1) A lightning regtest network for safe and rapid payment experimentation, 2) An IPFS service for real-time communication and 3) A Feedback Engine that facilitates RAG functionalities, simplified to operate over a basic system event log and maybe integrated with third-party RAG services. This configuration is optimized for isolating, testing, and developing agent behaviors within a private network, allowing for detailed simulation of peer interactions and economic behaviors without the risks of live network conditions 

\textbf{Core Feature: IPFS PubSub Communication}\\
\textit{Challenge}: How can SWE-Agents interact in real-time within the sandbox environment ?\\
\textit{Solution}: We simulate real-time IPFS PubSub communication over a Docker network by isolating each agent in its own container and leveraging IPFS PubSub for decentralized messaging. This allows agents to publish and subscribe to events in real time across the Docker network. By replicating real-world peer-to-peer communication without relying on a central server.

\textbf{Core Feature: Lightning Network Transactions}\\
\textit{Challenge}: How can SWE-Agents conduct instant cryptocurrency transactions within the sandbox environment?\\
\textit{Solution}: We simulate Lightning Network transactions in our sandbox environment by using Polar \cite{lightningpolar}, a tool that simplifies the creation and management of local Lightning Network clusters. These clusters run on the regtest network, a specialized Bitcoin testing network that allows for rapid, customizable testing without the risks or delays of real-world Bitcoin transactions. This setup enables us to simulate secure, fast micropayments between agents within the sandbox environment without the complexities of a live network.

\begin{figure*}[tbp]
    \centering
    \includegraphics[width=0.75\linewidth]{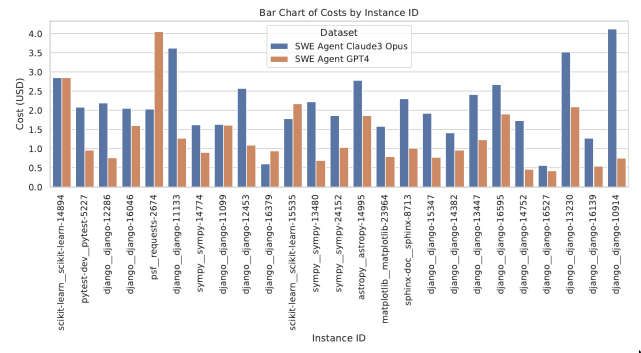} 
    \caption{Costs of SWE-agent solving different issues with Claude3 Opus and GPT-4}
    \label{fig:eda}
\end{figure*}

\textbf{Core Feature: Built-in RAG System}\\
\textit{Challenge}: How can SWE-Agents conduct inquiries about their environment?\\
\textit{Solution}: We simulate the environment’s built-in Retrieval-Augmented Generation (RAG) system for each SWE-agent by enabling access to a query engine that leverages a combination of open-source models (e.g., LLaMA 3.2) and third-party models like OpenAI’s GPT-4o within the agents' Docker containers. This setup allows the RAG system to dynamically retrieve relevant, real-time data from system event logs, providing each agent with up-to-date environment feedback for more effective decision-making.

The \texttt{ghissuemarket} CLI serves as a  wrapper that integrates  internal tools—like \texttt{ghissuemarket\--feedback\_engine} for Retrieval-Augmented Generation (RAG)—and external utilities like \texttt{lncli} for managing payments and \texttt{ipfs pubsub} for communication. This design adheres to the Unix philosophy, where each tool performs a well-defined task and can be combined with others  through the CLI to efficiently perform complex operations. This toolbox design—comprising \texttt{ghissuemarket}, \texttt{lncli}, and \texttt{ipfs pubsub}—demonstrates how independent tools, each serving a specific function, can be integrated to form a cohesive system. 

By leveraging this architecture, agents can manage decentralized auctions, execute payment transactions, and facilitate peer-to-peer communication without needing to handle the intricacies of each individual tool’s configuration and through this unified interface, agents can focus on their core activities, thereby enhancing the overall efficiency and flexibility of the experimental sandbox environment.

\section{FUTURE PLANS}
In this paper, we have presented  GHIssueMarket, a novel intelligent software engineering economics sandbox environment that supports the economic experimentation of SWE-Agents, simulating an envisioned peer-to-peer multi-agent system for GitHub issues outsourcing auctions. 

We plan to conduct an extensive series of experiments within the GHIssueMarket sandbox, with each experiment carefully designed to explore a core hypothesis:

1) We hypothesize that SWE-Agents can exhibit enhanced economic viability by resolving issues more cost-effectively and optimizing individual budgets compared to a baseline scenario without an outsourcing environment. This hypothesis is grounded on the insights from our exploratory data analysis, presented in Figure \ref{fig:eda}. This analysis of SWE-bench \cite{jimenez2024swebench} data serves as an exemplary case, revealing differences in operational costs between the two agent models across various tasks. Notably, SWE Agent Claude3 Opus consistently incurs higher costs than SWE Agent GPT4, highlighting opportunities for improving cost-efficiency through bidding and auctioning strategies. 

2) We hypothesize that increased competition among SWE-Agents within the outsourcing environment will drive down the average cost of issue resolution, leading to more efficient resource use compared to a non-competitive baseline.

3) We hypothesize that agents may opt for specialization in domain-specific tasks, which could enhance their comparative advantage and cost-effectiveness in those domain-specific issues. 

4) We hypothesize that agents will exhibit adaptive behavior in response to interactions with human participants by adjusting their strategies based on observed human bidding patterns and decision-making behaviors, where such adaptive behaviors would generally depend on the agents' economic rationality.

Nonetheless, these experiments may uncover SWE-Agents' undesirable behaviors that may need to be addressed and refined before they can be safely deployed in real-world applications. Our primary goal, however, is to analyze and understand the outcomes of these experiments, leveraging insights from a range of established fields such as Multi-Agent Systems (MAS), Game Theory, Mechanism Design, Agent-Based Computational Economics (ACE), and Generative Agent-Based Modeling (GABMs). By drawing from these interdisciplinary perspectives, we aim to gain a more comprehensive understanding of SWE-Agents' economic viability, with the goal of either improving the agents directly or enhancing their environments to better support their performance.

We open-source our software artifacts at \url{https://github.com/lascam-UFU/ghissuemarketsandbox}, inviting contributions and encouraging researchers and practitioners to use the GHIssueMarket sandbox for their SWE-Agents' economic experimentation. The insights gained from these experiments will contribute to the emerging field of Intelligent Software Engineering Economics (ISEE).

\bibliographystyle{plain}
\bibliography{ref}

\begin{thebibliography}{10}

\bibitem{auctionarena}
Aucarena: An auction-based evaluation suite for large language models.
\newblock \url{https://auction-arena.github.io/}.
\newblock Accessed: 2024-10-10.

\bibitem{lightningpolar}
Lightning polar - easy setup and management of lightning network nodes.
\newblock \url{https://lightningpolar.com/}, 2024.
\newblock Accessed: 2024-10-11.

\bibitem{bacon2010software}
David~F. Bacon et~al.
\newblock Software economies.
\newblock In {\em Proc. of the FSE/SDP Workshop on Future of Software Engineering Research (FoSER)}, pages 7--12. ACM, 2010.

\bibitem{boehm1981}
B.~W. Boehm.
\newblock {\em Software Engineering Economics}.
\newblock Prentice-Hall, 1981.

\bibitem{cognition_devin}
{Cognition AI}.
\newblock {Introducing Devin: The Next Generation AI for Software Development}.
\newblock \url{https://www.cognition.ai/blog/introducing-devin}, 2024.
\newblock Accessed: 2024-10-10.

\bibitem{dakhel2023githubcopilotaipair}
Arghavan~Dakhel et~al.
\newblock {GitHub Copilot AI pair programmer: Asset or Liability?}
\newblock \url{https://arxiv.org/abs/2206.15331}, 2023.

\bibitem{Zhang_2023}
Beiqi~Zhang et~al.
\newblock Practices and challenges of using {GitHub Copilot}: An empirical study.
\newblock In {\em Proc. of the 35th International Conference on Software Engineering and Knowledge Engineering}, volume 2023 of {\em SEKE2023}, page 124–129. KSI Research Inc., July 2023.

\bibitem{qian-etal-2024-chatdev}
Chen~Qian et~al.
\newblock {C}hat{D}ev: Communicative agents for software development.
\newblock In Lun-Wei Ku, Andre Martins, and Vivek Srikumar, editors, {\em Proc. of the 62nd Annual Meeting of the Association for Computational Linguistics (Volume 1)}, pages 15174--15186. ACL, August 2024.

\bibitem{yang2024sweagent}
John~Yang et~al.
\newblock {SWE-agent: Agent-Computer Interfaces Enable Automated Software Engineering}.
\newblock arXiv:2405.15793, 2024.

\bibitem{chen2021evaluatinglargelanguagemodels}
Mark~Chen et~al.
\newblock Evaluating large language models trained on code, 2021.

\bibitem{lewis2021retrievalaugmentedgenerationknowledgeintensivenlp}
Patrick~Lewis et~al.
\newblock Retrieval-augmented generation for knowledge-intensive {NLP} tasks.
\newblock \url{https://arxiv.org/abs/2005.11401}, 2021.

\bibitem{wu2024autogen}
Qingyun~Wu et~al.
\newblock {AutoGen}: Enabling next-gen {LLM} applications via multi-agent conversation framework.
\newblock In {\em COLM}, 2024.

\bibitem{10.5555/3495724.3495883}
Tom B.~Brown et~al.
\newblock Language models are few-shot learners.
\newblock In {\em Proc. of the 34th International Conference on Neural Information Processing Systems}, NIPS '20, Red Hook, NY, USA, 2020. Curran Associates Inc.

\bibitem{ghaffarzadegan2023generative}
Navid Ghaffarzadegan et~al.
\newblock Generative agent-based modeling: Unveiling social system dynamics through coupling mechanistic models with generative artificial intelligence.
\newblock {\em System Dynamics Review}, September 2023.

\bibitem{github_copilot}
GitHub.
\newblock {GitHub Copilot}.
\newblock \url{https://github.com/features/copilot}, 2024.
\newblock Accessed: 2024-10-10.

\bibitem{hurwicz2006designing}
Leonid Hurwicz and Stanley Reiter.
\newblock {\em Designing Economic Mechanisms}.
\newblock Cambridge University Press, Cambridge, UK, 2006.

\bibitem{jimenez2024swebench}
Carlos~E Jimenez, John Yang, Alexander Wettig, Shunyu Yao, Kexin Pei, Ofir Press, and Karthik~R Narasimhan.
\newblock {SWE}-bench: Can language models resolve real-world {Github} issues?
\newblock In {\em The Twelfth International Conference on Learning Representations}, 2024.

\bibitem{10.5555/3600270.3601883}
Takeshi Kojima, Shixiang~Shane Gu, Machel Reid, Yutaka Matsuo, and Yusuke Iwasawa.
\newblock Large language models are zero-shot reasoners.
\newblock In {\em Proc. of the 36th International Conference on Neural Information Processing Systems}, NIPS '22, Red Hook, NY, USA, 2024. Curran Associates Inc.

\bibitem{martinazzi2020evolving}
Stefano Martinazzi and Alessandra Flori.
\newblock The evolving topology of the lightning network: Centralization, efficiency, robustness, synchronization, and anonymity.
\newblock {\em PLoS ONE}, 15(1):e0225966, 2020.

\bibitem{mcconnell2006software}
Steve McConnell.
\newblock {\em Software Estimation: Demystifying the Black Art}.
\newblock Microsoft Press, Redmond, WA, 2006.

\bibitem{naur1969software}
Peter Naur and Brian Randell, editors.
\newblock {\em Software Engineering: Report on a Conference Sponsored by the NATO Science Committee, Garmisch, Germany, 7-11 October 1968}.
\newblock NATO Scientific Affairs Division, Brussels, Belgium, 1969.

\bibitem{osborne1994course}
Martin~J. Osborne and Ariel Rubinstein.
\newblock {\em A Course in Game Theory}.
\newblock MIT Press, Cambridge, MA, 1994.

\bibitem{10.1145/2744200}
Sebastian Proksch, Johannes Lerch, and Mira Mezini.
\newblock Intelligent code completion with {Bayesian} networks.
\newblock {\em ACM Trans. Softw. Eng. Methodol.}, 25(1), December 2015.

\bibitem{pmlr-v235-raman24b}
Narun~Krishnamurthi Raman, Taylor Lundy, Samuel~Joseph Amouyal, Yoav Levine, Kevin Leyton-Brown, and Moshe Tennenholtz.
\newblock {STEER}: Assessing the economic rationality of large language models.
\newblock In {\em Proceedings of the 41st International Conference on Machine Learning}, volume 235, pages 42026--42047. PMLR, 2024.

\bibitem{richiardi2012agent}
Matteo~G. Richiardi.
\newblock Agent-based computational economics: A short introduction.
\newblock {\em The Knowledge Engineering Review}, 27(2):137--149, 2012.

\bibitem{shoham2008multiagent}
Yoav Shoham and Kevin Leyton-Brown.
\newblock {\em Multiagent Systems: Algorithmic, Game-Theoretic, and Logical Foundations}.
\newblock Cambridge University Press, New York, NY, USA, 2008.

\bibitem{ipfs_pubsub}
IPFS Team.
\newblock Ipfs \#25: Pubsub, July 2017.
\newblock Accessed: 2024-10-10. Available at: \url{https://blog.ipfs.tech/25-pubsub/}.

\bibitem{10.1145/3570220}
Matt Welsh.
\newblock The end of programming.
\newblock {\em Commun. ACM}, 66(1):34–35, December 2022.

\bibitem{7390282}
W.~Eric Wong, Ruizhi Gao, Yihao Li, Rui Abreu, and Franz Wotawa.
\newblock A survey on software fault localization.
\newblock {\em IEEE Transactions on Software Engineering}, 42(8):707--740, 2016.

\bibitem{10.1145/3172871.3172891}
Tao Xie.
\newblock Intelligent software engineering: Synergy between {AI and Software Engineering}.
\newblock In {\em Proc. of the 11th Innovations in Software Engineering Conference}, ISEC '18, New York, NY, USA, 2018. Association for Computing Machinery.

\bibitem{10.1145/3631974}
Quanjun Zhang, Chunrong Fang, Yuxiang Ma, Weisong Sun, and Zhenyu Chen.
\newblock A survey of learning-based automated program repair.
\newblock {\em ACM Trans. Softw. Eng. Methodol.}, 33(2), December 2023.

\end{thebibliography}

\balance

\end{document}